\documentclass[10pt,conference]{IEEEtran}

\usepackage{amsfonts}
\usepackage{amssymb,epic,eepic}
\usepackage{amsmath}
\usepackage{dcolumn}
\usepackage{bm}

\newtheorem{theorem}    {Theorem}       [section]
\newtheorem{lemma}      {Lemma}         [section]

\newcommand{\hr}{{\mathcal H}}
\newcommand{\cn}{{\mathcal N }}
\newcommand{\E}{{\mathcal E }}

\newcommand{\cs}{{\mathcal S}}
\newcommand{\crr}{{\mathcal R}}
\newcommand{\fr}{{\mathcal F}}
\newcommand{\gr}{{\mathcal G}}
\newcommand{\fri}{{\mathfrak I}}
\newcommand{\kr}{{\mathcal K}}

\newcommand{\cc}{{\mathbb C}}
\newcommand{\rr}{{\mathbb R}}
\newcommand{\nn}{{\mathbb N}}
\newcommand{\idn}{\mathbf{1}}

\newcommand{\vphi}{{\varphi}}           

\newcommand{\A}{\mathcal A}

\newcommand{\D}{\mathcal D}
\newcommand{\U}{\mathcal U}

\newcommand{\tr}{\mathrm{tr}}

\begin{document}

\title{Entanglement Transmission Capacity of Compound Channels}
\author{
\authorblockN{Igor Bjelakovi\'c}
\authorblockA{Heinrich-Hertz-Lehrstuhl\\
f\"ur Mobilkommunikation (HFT 6)\\
and Institut f\"ur Mathematik\\
Technische Universit\"at Berlin, Germany\\
Email: igor.bjelakovic@mk.tu-berlin.de}
\and
\authorblockN{Holger Boche}
\authorblockA{Heinrich-Hertz-Lehrstuhl\\
f\"ur Mobilkommunikation (HFT 6)\\
and Institut f\"ur Mathematik\\
Technische Universit\"at Berlin, Germany
\\
Email: holger.boche@mk.tu-berlin.de}
\and
\authorblockN{Janis N\"otzel
}
\authorblockA{Heinrich-Hertz-Lehrstuhl\\
f\"ur Mobilkommunikation (HFT 6)\\
Technische Universit\"at Berlin, Germany\\
Email: janis.noetzel@mk.tu-berlin.de}
}
\maketitle

\begin{abstract}
We determine the optimal achievable rate at which entanglement can be reliably transmitted when the memoryless channel used during transmission is unknown both to sender and receiver. To be more precise, we assume that both of them only know that the channel belongs to a given set of channels. Thus, they have to use encoding and decoding schemes that work well for the whole set.
\end{abstract}

\section{Introduction}
One of the main goals of quantum Shannon theory is the
determination of optimal transmission rates for various quantum
communication tasks. In contrast to classical information theory to every
quantum channel we can associate various capacities each of which
characterizes the optimal rates in a specific communication scenario.
In this paper we focus on the determination of the entanglement transmission capacity of quantum  compound channels.\\
The correct formula describing this capacity for a single channel has been identified in \cite{bkn,devetak,shor}.
Of particular interest for our work are the later on developments by Klesse \cite{klesse} and Hayden, Horodecki, Winter and Yard \cite{hayden} which are based on a decoupling idea that can be traced back to Schumacher and Westmoreland \cite{schumacher-westmoreland}.\\
We use their approach to determine the optimal achievable entanglement transmission rate under channel uncertainty: while sustaining the assumption of memoryless communication, we assume that sender as well as receiver only know that the channel they use belongs to some given set of channels. This describes a somewhat more realistic situation since exact channel knowledge will hardly ever be given in applications.\\
Due to space limitation we will only give the proof of the direct part of the
coding theorem for finite compound channels. The extension to the general
case, the proof of the converse part and the relation to the entanglement-generating capacity of compound channels can be picked up
in the accompanying paper \cite{bbn-2}.\\
The paper is organized as follows: We first fix the notation in section \ref{sec:Notation and Conventions}. In section \ref{sec:Codes and Capacity} we introduce our model and state the main theorem. Section \ref{sec:One-Shot Results} contains two results concerning existence of recovery operations of a certain performance and behavior of entanglement fidelity under disturbance of a channel through a projection. The proof of our main theorem further uses some basic properties of typical projections and operations, which are stated in section \ref{sec:Typical Projections and Operations}. From there we pass on to the proof of our main theorem in Section \ref{sec:Proof of Theorem}.
\section{Notation and Conventions\label{sec:Notation and Conventions}}
All Hilbert spaces are assumed to have finite dimension and are over the field $\mathbb C$. $\mathcal{S}(\hr)$ is the set of states, i.e. positive semi-definite operators with trace $1$ acting on the Hilbert space $\hr$. Pure states are given by projections onto one-dimensional subspaces. A vector of unit length spanning such a subspace will therefore be referred to as a state vector.\\
The set of completely positive trace preserving (CPTP) maps
between the operator spaces $\mathcal{B}(\hr)$ and
$\mathcal{B}(\kr)$ is denoted by $\mathcal{C}(\hr,\kr)$.
$\mathcal{C}^{\downarrow}(\hr,\kr)$ stands for the set of
completely positive trace decreasing maps between
$\mathcal{B}(\hr)$ and $\mathcal{B}(\kr)$. $\mathfrak{U}(\hr)$ will denote in what follows the group of unitary operators acting on $\hr$. For a Hilbert space $\gr\subset \hr$ we will always identify $\mathfrak{U}(\gr)$ with a subgroup of $\mathfrak{U}(\hr)$ in the canonical way. For any projection $q\in\mathcal{B}(\hr)$ we set $q^\perp:=\idn_{\hr}-q$. Each projection $q\in\mathcal{B}(\hr)$ defines a completely positive trace decreasing map $\mathcal{Q}$ given by $\mathcal{Q}(a):=qaq$ for all $a\in\mathcal{B}(\hr)$. In a similar fashion any $u\in \mathfrak{U}(\hr)$ defines a $\U\in \mathcal{C}(\hr,\hr)$ by $\U(a):=uau^{\ast}$ for $a\in\mathcal{B}(\hr)$.\\
We use the base two
logarithm which is denoted by $\log$. The von Neumann entropy of
a state $\rho\in\mathcal{S}(\hr)$ is given by $S(\rho):=-\textrm{tr}(\rho \log\rho)$.
The coherent information for $\cn\in \mathcal{C}(\hr,\kr) $ and
$\rho\in\mathcal{S}(\hr)$ is defined by $I_c(\rho, \cn):=S(\cn (\rho))- S( (id_{\hr}\otimes \cn)(|\psi\rangle\langle \psi|)  )$,
where $\psi\in\hr\otimes \hr$ is an arbitrary purification of the state $\rho$. Following the usual conventions we let $S_e(\rho,\cn):=S( (id_{\hr}\otimes \cn)(|\psi\rangle\langle \psi|)  )$ denote the entropy exchange.\\
For $\rho\in\mathcal{S}(\hr)$ and $\cn\in
\mathcal{C}^{\downarrow}(\hr,\kr)$ the entanglement Fidelity is given by $F_e(\rho,\cn):=\langle\psi, (id_{\hr}\otimes \cn)(|\psi\rangle\langle \psi|)     \psi\rangle$, with $\psi\in\hr\otimes \hr$ being an arbitrary purification of the state $\rho$.\\
In the following, a compound channel is identified with the set $\fri\subset\mathcal C(\hr,\kr)$ of its constituents. It is called finite if $\fri$ consists of finitely many elements.
\section{Codes, Capacity and Main Result\label{sec:Codes and Capacity}}
An $(l,k_l)-$\emph{entanglement transmission code} for the compound channel $\fri$ is a pair $(\mathcal{P}^l, \crr^l)$ of CPTP maps $\mathcal{P}^l\in\mathcal{C}(\fr_l,\hr^{\otimes l})  $ where $\fr_l$ is a Hilbert space with $k_l=\dim \fr_l$ and $\crr^l\in\mathcal{C}(\kr^{\otimes l},\fr_l')$ with $\fr_l\subset \fr_l'$.\\
A nonnegative number $R$ is called an achievable rate for (entanglement transmission through)
$\fri$ if there is a sequence of
$(l,k_l)$-entanglement transmission codes such that
\begin{enumerate}
\item $\liminf_{l\to\infty}\frac{1}{l}\log k_l\ge R$, and \item
$\lim_{l\to\infty}\inf_{\cn\in\fri}F_e(\pi_{\fr_l},\crr^l\circ \cn^{\otimes l}\circ\mathcal{P}^l)=1 $.
\end{enumerate}
The \emph{entanglement transmission capacity} $Q(\fri)$ of the compound
channel $\fri$ is given by
\begin{equation*}
 Q(\fri):=\sup\{R\in\rr_{+}: R \textrm{ is achievable for } \fri   \}.
\end{equation*}
Our main result can now be formulated as follows:
\begin{theorem}\label{main-theorem}
Let $\fri\subset\mathcal C(\hr,\kr)$ be a compound channel. The entanglement transmission capacity of $\fri$ is given by
$$Q(\fri)=\lim_{l\rightarrow\infty}\frac{1}{l}\max_{\rho\in\mathcal S(\hr^{\otimes l})}\inf_{\cn\in\fri}I_c(\rho,\cn^{\otimes l}).$$
\end{theorem}
\emph{\underline{Remark}. Corresponding results for the entanglement transmission capacity of a compound channel with informed encoder or informed decoder can be found in \cite{bbn-2}. It is a remarkable fact that the proof of the coding theorem for an informed encoder is not, as in the classical case, just a trivial modification of the one for Theorem \ref{main-theorem}.}
\section{One-Shot Results\label{sec:One-Shot Results}}
This section contains essentially two statements. The first gives an estimate on the performance of universal recovery operations for a given finite set of channels. The second relates the entanglement fidelity of a coding-decoding procedure to that of a disturbed version of the procedure, where disturbance means application of a projection after using the channel.\\
Both results give rather loose bounds that become sharp enough only in the asymptotic limit.
\subsection{Performance of Recovery Operations}
Before we turn our attention to quantum compound channels we will shortly describe a part of recent developments in coding theory for single (i.e. perfectly known) channels as given in \cite{klesse} and \cite{hayden}. Both approaches are based on a decoupling idea which is closely related to approximate error correction. In order to state this decoupling lemma we need some notational preparation.\\
Let $\rho\in\mathcal{S}(\hr)$ be given and consider any purification $\psi\in\hr_a\otimes \hr$, $\hr_a=\hr$, of $\rho$. According to Stinespring's representation theorem any $\cn\in\mathcal{C}^{\downarrow}(\hr,\kr)$ is given by
\begin{equation}\label{stinespring-1}
  \cn (\ \cdot\ )=\textrm{tr}_{\hr_e}((\idn_{\hr}\otimes p_e)v(\ \cdot\ )v^{\ast} ),
\end{equation}
where $\hr_e$ is a suitable finite-dimensional Hilbert space, $p_e$ is a projection onto a subspace of $\hr_e$, and $v:\hr \to\kr\otimes \hr_e $ is an isometry. \\
Let us define a pure state on $\hr_a\otimes \kr\otimes \hr_e$ by the formula
\[  \psi':= \frac{1}{\sqrt{\textrm{tr}(\cn(\rho))}}(\idn_{\hr_a\otimes \kr}\otimes p_e)(\idn_{\hr_a}\otimes v) \psi.  \]
We set
\[\rho':=\textrm{tr}_{\hr_a\otimes \hr_e}(|\psi'\rangle\langle \psi'|),\quad \rho'_{ae}:= \textrm{tr}_{\kr}(|\psi'\rangle\langle \psi'|), \]
\[\rho_a:=\textrm{tr}_{\kr\otimes\hr_e}(|\psi'\rangle\langle \psi'|),\quad \rho'_e:=\textrm{tr}_{\hr_a\otimes \kr}(|\psi'\rangle\langle \psi'|).  \]
The announced decoupling lemma can now be stated as follows.
\begin{lemma}[Cf. \cite{klesse, hayden}]\label{decoupling-lemma}
For any $\cn\in \mathcal{C}^{\downarrow}(\hr,\kr) $ there exists a recovery operation $\crr \in \mathcal{C}(\kr,\hr) $ with
\[F_e(\rho, \crr\circ \cn)\ge w-||w\rho'_{ae}-w\rho_a\otimes \rho'_e||_1,  \]
where $w=\textrm{tr}(\cn(\rho))$.
\end{lemma}
We will make use of this lemma in the proof of the following
theorem, which is the heart of the proof of Theorem \ref{main-theorem}. In order to state the theorem, we need to introduce the code entanglement fidelity which is, for $\rho\in\mathcal S(\hr),\cn\in\mathcal
C(\hr,\kr)$ (referring to $\rho$ as the code) given by
$$F_{c,e}(\rho,\cn):=\max_{\crr\in\mathcal C(\kr,\hr)}F_{e}(\rho,\crr\circ\cn).$$
\begin{theorem}[One-Shot Result for Averaged Channel]\label{convex-klesse}
Let the Hilbert space $\hr$ be given and consider subspaces
$\E\subset\gr\subset \hr$ with $\dim \E=k$. For any choice of
$\cn_1,\ldots,\cn_N\in \mathcal{C}^{\downarrow}(\hr,\kr) $ each
allowing a representation with $n_j$ Kraus operators, $j=1,\ldots
, N$, and and for any $u\in\mathfrak{U}(\gr)$ we set
\[\cn:=\frac{1}{N}\sum_{j=1}^{N}\cn_j,\ \ \ \ \cn_u:= \frac{1}{N}\sum_{j=1}^{N}\cn_j\circ \U.\ \ \ \mathrm{Then}  \]
\begin{eqnarray*}
 \int_{\mathfrak{U}(\gr)}F_{c,e}(\pi_{\E},\cn_u)du\geq \textrm{tr}(\cn(\pi_{\gr}))
- 2 \sum_{j=1}^N \sqrt{k n_j }||\cn_j (\pi_{\gr})||_2,
\end{eqnarray*}
where the integration is with respect to the normalized Haar measure on $\mathfrak{U}(\gr)$ and $\pi_\E,\pi_\gr$ are the maximally mixed states on $\E$ and $\gr$.
\end{theorem}
\emph{\underline{Remark}. The above Theorem gives a lower bound on the code entanglement fidelity of an averaged channel. Since entanglement fidelity is affine in the operation, $F_e(\pi_{\fr_l},\frac{1}{N}\sum_{i=1}^N\cn_i^{\otimes l}\circ\mathcal{P}^l)\geq1-\epsilon_l$ implies $F_e(\pi_{\fr_l},\cn_i^{\otimes l}\circ\mathcal{P}^l)\geq1-N\epsilon_l$ for every $i\in\{1,\ldots,N\}$. If $\fri$ is finite and $\epsilon_l$ becomes arbitrarily small for good codes, this Theorem gives a sufficient estimate. The case of general $\fri$ exploits the difference in polynomial growth of the number $N_l$ of approximating channels for $\fri$ versus exponential decay of $\epsilon_l$.}\\
For the proof of this Theorem, we shall need the following two lemmata:
\begin{lemma}[Cf. \cite{bbn-1}]\label{matrix-lemma}
Let $L$ and $D$ be $N\times N$ matrices with non-negative entries which satisfy
\begin{equation}\label{matrix-1}
 L_{jl}\le L_{jj}, \quad L_{j l}\le L_{ll},\ \ \mathrm{and}\ \ D_{jl}\le \max \{ D_{jj}, D_{ll}\}
\end{equation}
for all $j,l\in \{ 1,\ldots, N \}$. Then
\[\sum_{j,l=1}^{N}\frac{1}{N}\sqrt{L_{jl}D_{jl}}\le 2\sum_{j=1}^{N}\sqrt{L_{jj}D_{jj}}.  \]
\end{lemma}
\begin{lemma}[Cf. \cite{bbn-2}]\label{lemma-trace-average}
Let $\E$ and $\gr$ be subspaces of $\hr$ with
$\E\subset\gr\subset\hr$ where $k:=\dim \E$, $d_{\gr}:=\dim
\gr$. $p$ and $p_{\gr}$ will denote the orthogonal projections
onto $\E$ and $\gr$. For a Haar distributed random variable $U$
with values in $\mathfrak{U}(\gr)$ and $x,y\in\mathcal{B}(\hr)$ we
define a random sesquilinear form
\[b_{UpU^{\ast}}(x,y):= \textrm{tr}(UpU^{\ast} x^{\ast}UpU^{\ast}y )-\frac{1}{k}\textrm{tr}(UpU^{\ast}x^{\ast})\textrm{tr}(UpU^{\ast}y).  \]
Then
\begin{eqnarray}
  \mathbb{E}\{b_{UpU^{\ast}}(x,y) \}&=& \frac{k^2-1}{d^2-1}\textrm{tr}(p_{\gr}x^{\ast}p_{\gr}y)\nonumber\\
  &&+\frac{1-k^2}{d(d^2-1 )}\textrm{tr}(p_{\gr}x^{\ast})\textrm{tr}(p_{\gr}y).\nonumber
\end{eqnarray}
\end{lemma}
\emph{Proof of Theorem \ref{convex-klesse}:}
We can assume without loss of generality that the numbering of the channels is chosen in such a way that $n_1\le n_2\le \ldots \le n_N$ holds for the numbers of Kraus operators of the maps $\cn_1,\ldots, \cn_N$. From Lemma \ref{decoupling-lemma} we know that for every $u\in\mathfrak U(\gr)$ there is a recovery operation $\crr$ such that
\begin{equation}\label{eq:convex-klesse-1}
 F_e(\pi_{\E}, \crr\circ \cn_u)\ge w-||w\rho'_{ae}-w\rho_a\otimes \rho'_e||_1,
\end{equation}
where we have used the notation introduced in the paragraph preceding Lemma \ref{decoupling-lemma} and the states on the RHS of equation (\ref{eq:convex-klesse-1}) now depend on $u$.\\
For each $j\in\{1,\ldots ,N\}$ let $\{b_{j,i}\}_{i=1}^{n_j}$ be the set of Kraus operators of $\cn_j$. Then $\cn_j\circ\U$ has Kraus operators $\{a_{j,i}\}_{i=1}^{n_j}$ given by $a_{j,i}=b_{j,i}u$. Let $\{f_1,\ldots, f_N  \}$ and $\{e_1,\ldots ,e_{n_N}  \}$ be arbitrary orthonormal bases of $\cc^{N}$ and $\cc^{n_N}$ with only imposed restriction that $e_1\otimes f_1=\psi_e$. Let the projection $p_e$ and unitary $v$ in (\ref{stinespring-1}) be chosen in such a way that for each $\phi\in \hr$ the relation
\begin{equation}\label{stinespring-2}
(\idn_{\hr}\otimes p_e)v (\phi\otimes e_1\otimes f_1)=\sum_{j=1}^{N}\sum_{i=1}^{n_j}\frac{1}{\sqrt{N}}(b_{j,i}\phi)\otimes e_i\otimes f_j,
\end{equation}
holds. For a purification $\psi\in \hr_a\otimes \hr$ of the state $\pi_{\E}$ we consider a Schmidt representation
\[\psi=\frac{1}{\sqrt{k}}\sum_{m=1}^{k}h_m\otimes g_m,  \]
with suitable orthonormal systems $\{ h_1,\ldots ,h_k \}$ and $\{ g_1,\ldots, g_k  \}$.\\
We use this representation to derive explicit representations of the states $\rho'_{ae},\rho_a,\rho'_e$ in terms of the Kraus operators of the operations $\cn_i$ and insert them into (\ref{eq:convex-klesse-1}). If we perform the unitary conjugation induced by the unitary map $x_{s,i,j}=h_s\otimes e_i\otimes f_j\mapsto x'_{s,i,j}= g_s\otimes e_i\otimes f_j $ followed by the complex conjugation of the matrix elements with respect to the matrix units $\{|x'_{s,i,j}\rangle\langle x'_{t,k,l}|  \}_{s,i,j,t,k,l }$ we obtain an anti-linear isometry $I$ with respect to the metrics induced by the trace distances on the operator spaces under consideration.\\
A calculation identical to that performed by Klesse \cite{klesse} and additionally using the triangle inequality for $||\cdot||_1$ as well as the relation $||a ||_1\le \sqrt{d}||a ||_2$, $d$ being the number of non-zero singular values of the operator $a$ shows that
\begin{eqnarray}
F_{c,e}(\pi_\E,\cn_u)\geq \tr(\cn_u(\pi_{\E}))-\sum_{j,l=1}^N\frac{1}{N}\sqrt{\frac{1}{k}L_{jl}D_{jl}(u)},\label{eq:convex-klesse-10}
\end{eqnarray}
where
$$D_{jl}(u):=
\sum_{i=1,r=1}^{n_j,n_l}(\textrm{tr}(p (a_{j,i}^{\ast}a_{l,r})^{\ast}p a_{j,i}^{\ast}a_{l,r})-\frac{1}{k}|\textrm{tr}(pa_{j,i}^{\ast}a_{l,r} )|^2)$$
(dependence on $u$ is through $a_{i,j}=b_{i,j}u$) and $L_{jl}:=\min\{n_j,n_l\}$.\\
Let $U$ be a random variable taking values in $\mathfrak{U}(\gr)$ according to the Haar measure of $\mathfrak{U}(\gr) $. Then we can infer from (\ref{eq:convex-klesse-10}) that
\begin{eqnarray}
\mathbb E F_{c,e}(\pi_\E,\cn\circ\U)&\geq&\mathbb E\tr(\cn\circ\U(\pi_{\E}))\label{eq:convex-klesse-11}\\
&&-\sum_{j,l=1}^N\frac{1}{N}\sqrt{\frac{1}{k}L_{jl}\mathbb E(D_{jl}(U))},\nonumber
\end{eqnarray}
where we have used concavity of the function $\sqrt{\ \cdot \ }$
and Jensen's inequality. Now, setting $D_{jl}:=\langle\cn_j(\pi_{\gr}),\cn_l(\pi_{\gr})\rangle_{HS}$, where $\langle \ \cdot \ ,\ \cdot \ \rangle_{HS} $ denotes the Hilbert-Schmidt inner product, and using Lemma \ref{lemma-trace-average} we obtain
\begin{eqnarray}\label{eq:convex-klesse-12}
\mathbb E D_{jl}(U)\le \textrm{tr}(\cn_j(\pi_{\gr})\cn_l(\pi_{\gr}) )=D_{jl}.
\end{eqnarray}
It is obvious that $L_{jl}\le L_{jj}$ and $L_{j l}\le L_{ll}$ hold. Moreover, the Cauchy-Schwarz inequality for the Hilbert-Schmidt inner product justifies the inequality $D_{jl}\le\max\{D_{jj}, D_{ll}\}$.\\
Therefore, an application of Lemma \ref{matrix-lemma} allows us to conclude from (\ref{eq:convex-klesse-11}) that
\begin{eqnarray*}
\mathbb{E} (F_{c,e}(\pi_{\E},\cn\circ\U))\ge \textrm{tr}(\cn (\pi_{\gr})) -2\sum_{j=1}^{N}\sqrt{kn_j}||\cn_j(\pi_{\gr})||_2,
\end{eqnarray*}
which is what we aimed to prove. $\Box$
\subsection{Projections and Entanglement Fidelity\label{sec:Projections and Entanglement Fidelity}}
\begin{lemma}\label{gentle-operator-lemma-for-F_e}
Let $\rho\in\mathcal S(\hr)$ for some Hilbert space $\hr$. Let, for some other Hilbert space $\kr$, $\A\in \mathcal C(\hr,\kr),\ \D\in \mathcal C(\kr,\hr)$, $q\in\mathcal B(\kr)$ be an orthogonal projection.
If for some $\epsilon>0$ the relation $F_e(\rho,\D\circ\mathcal{Q}\circ\A)\geq1-\epsilon $ holds, then
\begin{equation}\label{eq:F_e-lemma-2}
F_e(\rho,\D\circ\A)\geq 1-3\epsilon.
\end{equation}
\end{lemma}
The following Lemma \ref{choi-lemma} contains an inequality which will be needed in the proof of Lemma \ref{gentle-operator-lemma-for-F_e}.
\begin{lemma}[Cf. \cite{bbn-2}]\label{choi-lemma}
Let $\D\in\mathcal C(\kr,\hr)$ and $x_1\perp x_2$, $z$
be state vectors, $x_1,x_2\in\kr,\ z\in\hr$. Then
\begin{equation}\label{eq:choi-lemma-1}
|\langle z,\D(|x_1\rangle\langle x_2|)z\rangle|\leq\sqrt{|\langle z,\D(\mathcal P_{x_1})z\rangle|\cdot|\langle z,\D(\mathcal P_{x_2})z\rangle|}\nonumber,
\end{equation}
where $\mathcal P_{y}:=|y\rangle\langle y|$ for arbitrary state vectors $y\in\hr,\kr$.
\end{lemma}\ \\
\emph{Proof of Lemma \ref{gentle-operator-lemma-for-F_e}.} Let $\dim\hr=h,\ \dim\kr=\kappa$,
$|\psi\rangle\langle\psi|\in\hr_a\otimes\hr$ be a purification of
$\rho$ (w.l.o.g. $\hr_a=\hr$). Set
$\tilde\D:=id_{\hr_a}\otimes\D,\ \tilde\A:=id_{\hr_a}\otimes\A,\
\tilde q:=\idn_{\hr_a}\otimes q$ and, as usual, $\tilde
q^\perp$ the orthocomplement of $\tilde q$ within
$\hr_a\otimes\kr$. Obviously,
\begin{eqnarray}
&&F_e(\rho,\D\circ\A)=\nonumber\\
&=&\langle\psi,\tilde\D\circ\tilde\A(|\psi\rangle\langle\psi|)\psi\rangle\nonumber\\
&=&\langle\psi,\tilde\D(\tilde q \tilde\A(|\psi\rangle\langle\psi|)\tilde q)\psi\rangle+\langle\psi,\tilde\D(\tilde q^\perp\tilde\A(|\psi\rangle\langle\psi|)\tilde q^\perp)\psi\rangle\nonumber\\
&&+\langle\psi,\tilde\D(\tilde q\tilde\A(|\psi\rangle\langle\psi|)\tilde q^\perp)\psi\rangle+\langle\psi,\tilde\D(\tilde q^\perp\tilde\A(|\psi\rangle\langle\psi|)\tilde q)\psi\rangle\nonumber\\
&\geq&\langle\psi,\tilde\D(\tilde q \tilde\A(|\psi\rangle\langle\psi|)\tilde q)\psi\rangle-2|\langle\psi,\tilde\D(\tilde q\tilde\A(|\psi\rangle\langle\psi|)\tilde q^\perp)\psi\rangle|\nonumber\\
&=&F_e(\rho,\D\circ\mathcal
Q\circ\A)-2|\langle\psi,\tilde\D(\tilde
q\tilde\A(|\psi\rangle\langle\psi|)\tilde
q^\perp)\psi\rangle|\label{eq:Entanglement Fidelity-1}.
\end{eqnarray}
We establish a lower bound on the second term on the RHS of
(\ref{eq:Entanglement Fidelity-1}). Let
\[ \tilde\A(|\psi\rangle\langle\psi|)=\sum_{i=1}^{\kappa\cdot
h}\lambda_i|a_i\rangle\langle a_i|,\]
 where $\{a_1,\ldots,a_{\kappa\cdot
h}\}$ are assumed to form an orthonormal basis. Now every $a_i$
can be written as $a_i=\alpha_ix_i+\beta_iy_i$ where $x_i\in
\mathrm{supp}(\tilde q)$ and $y_i\in \mathrm{supp}(\tilde
q^\perp)$, $i\in\{1,...,\kappa\cdot h\}$, are state vectors and
$\alpha_i,\beta_i\in\mathbb C$. Define $\sigma:=\tilde
A(|\psi\rangle\langle\psi|)$, then
\begin{eqnarray}
\sigma&=&\sum_{j=1}^{\kappa\cdot h}\lambda_j(|\alpha_j|^2|x_j\rangle\langle x_j|+\alpha_j\beta_j^\ast|x_j\rangle\langle y_j|\nonumber\\
&&\qquad+\beta_j\alpha_j^\ast|y_j\rangle\langle x_j|+|\beta_j|^2|y_j\rangle\langle y_j|).\label{eq:Entanglement
Fidelity-2}
\end{eqnarray}
Set $X:=|\langle\psi,\tilde\D(\tilde
q\tilde\A(|\psi\rangle\langle\psi|)\tilde q^\perp)\psi\rangle|$.
Then, using the decomposition (\ref{eq:Entanglement Fidelity-2}) and the abbreviation $\mathcal P_{w}:=|w\rangle\langle w|$ (for $w\in\kr$ being a state-vector)
\begin{eqnarray}
X&=&|\langle\psi,\tilde\D(\tilde q\sigma q^\perp)\psi\rangle|\nonumber\\
&\leq&\sum_{i=1}^{\kappa\cdot h}|\lambda_i\alpha_i\beta_i^*|\cdot|\langle\psi,\tilde\D(|x_i\rangle\langle y_i|)\psi\rangle|\nonumber\\
&\overset{\mathbf a}{\leq}&\sum_{i=1}^{\kappa\cdot h}|\alpha_i\beta_i^*|\lambda_i\sqrt{|\langle\psi,\tilde\D(\mathcal P_{x_i})\psi\rangle\langle\psi,\tilde\D(\mathcal P_{y_i})\psi\rangle}|\nonumber\\
&\overset{\mathbf b}{\leq}&\sum_{i=1}^{\kappa\cdot
h}\lambda_i|\alpha_i|^2\langle\psi,\tilde\D(\mathcal P_{x_i})\psi\rangle\sum_{j=1}^{\kappa\cdot
h}\lambda_j|\beta_j|^2\langle\psi,\tilde\D(\mathcal P_{y_j})\psi\rangle.\nonumber\\
&=& F_e(\rho,\D\circ\mathcal Q\circ\A)\cdot F_e(\rho,\D\circ\mathcal Q^\perp\circ\A)\nonumber\\
&\overset{\mathbf c}{\leq}&\epsilon.\label{eq:Entanglement Fidelity-4}
\end{eqnarray}
Here, $\mathbf a$ follows from utilizing Lemma \ref{choi-lemma}, $\mathbf b$ is an application of the Cauchy-Schwarz inequality and $\mathbf c$ is true by assumption.\\
The inequality (\ref{eq:Entanglement Fidelity-4}) establishes (\ref{eq:F_e-lemma-2}). $\Box$
\section{Typical Projections and Operations\label{sec:Typical Projections and Operations}}
At this point, we introduce the minimal amount of statements about typical projections and operations that is needed for the proof of Theorem \ref{main-theorem}. The reader interested in more details is referred to \cite{bbn-1,bbn-2} and references therein. The basic idea is that we throw away some non-essential information about an object and get nice estimates in return.
\begin{lemma}\label{lemma-typical-1}
There is a real number $c>0$ such that for every two Hilbert spaces $\hr,\kr$ the following hold:\\
There are functions $h:\mathbb N\rightarrow\mathbb R_+$ and $\varphi:(0,1/2)\rightarrow\mathbb R_+$ with $h(l)\searrow0$ and $\varphi(\delta)\searrow0$ (Setting $d:=\dim\hr$, $\kappa:=\dim\kr$, $h$ and $\varphi$ are given by $h(l):=\frac{d\cdot\kappa}{l}\log(l+1)\ \forall l\in\nn$ and $\varphi(\delta):=-\delta\log\frac{\delta}{d\cdot\kappa}\ \forall\delta\in(0,1/2)$) such that\\
\textbf{A)} For any $\rho\in \cs (\hr),\ \delta\in(0,1/2),\
l\in\mathbb N$ there is an orthogonal projection $q_{\delta,l}\in
\mathcal{B}(\hr)^{\otimes l}$ called frequency-typical projection
that satisfies
\begin{enumerate}
\item $\textrm{tr}(\rho^{\otimes l}q_{\delta,l})\ge
1-2^{-l(c\delta^2-h(l))}$,
\item $q_{\delta,l}\rho^{\otimes l}q_{\delta,l}\le
2^{-l(S(\rho)-\varphi(\delta) )}q_{\delta,l}  $.
\end{enumerate}
The inequality 2) implies
\[||q_{\delta,l}\rho^{\otimes l}q_{\delta,l}||_2^2 \le 2^{-l(S(\rho)-\varphi(\delta))}. \]
\textbf{B)} For each $\cn\in
\mathcal{C}(\hr,\kr)$, $\delta\in (0,1/2)$, $l\in\nn$ and maximally mixed state $\pi_\gr$ on some
subspace $\gr\subset\hr$ there is an operation
$\cn_{\delta,l}\in\mathcal C^\downarrow(\hr^{\otimes
l},\kr^{\otimes l})$ called reduced operation with respect to
$\cn$ and $\pi_\gr$ that satisfies
\begin{enumerate}\setcounter{enumi}{2}
\item $\textrm{tr}(\cn_{\delta,l}(\pi_{\gr}^{\otimes l}))\ge
1-2^{-l(c\delta^2-h(l))}$,
\item $\cn_{\delta,l}$ has a Kraus representation with at
most $n_{\delta,l}\le 2^{l(S_{e}(\pi_{\gr},\cn)+\varphi(\delta)+h(l))}$
Kraus operators. \item For every state $\rho\in\mathcal
S(\hr^{\otimes l})$ and every two channels $\mathcal I\in\mathcal
C^\downarrow(\hr^{\otimes l},\hr^{\otimes l})$ and $\mathcal
L\in\mathcal C^\downarrow(\kr^{\otimes l},\hr^{\otimes l})$ the
inequality $F_e(\rho,\mathcal L\circ\cn_{\delta,l}\circ\mathcal
I)\leq F_e(\rho,\mathcal L\circ\cn^{\otimes l}\circ\mathcal I)$ is
fulfilled.
\end{enumerate}
\end{lemma}
\section{Proof of Theorem \ref{main-theorem}\label{sec:Proof of Theorem}}
We will restrict our proof to the case that $\fri$ consists of finitely many elements. Also, we only prove the direct part $Q(\fri)\geq\lim_{l\rightarrow\infty}\frac{1}{l}\inf_{\cn\in\fri}\max_{\rho\in\mathcal S(\hr^{\otimes l})}I_c(\rho,\cn^{\otimes l})$. The converse part for finite $\fri$ follows from an application of Lemma 6 in \cite{devetak}. In order to pass on to the case of general $\fri$ one approximates $\fri$ by a sequence $(\fri_l)_{l\in\nn}$ of finite compound channels. It has to be taken care that the numbers $N_l:=|\fri_l|$ increase subexponentially fast in $l$. All calculations are carried out in our papers \cite{bbn-2} and \cite{bbn-1}.\\
Let us consider a compound channel given by a finite set $\fri:=\{\cn_1,\ldots ,\cn_{N} \}\subset \mathcal{C}(\hr,\kr)$ and a subspace $\gr\subset\hr$. For every $l\in\mathbb N$, we choose a subspace $\fr_l\subset\gr^{\otimes l}$. As usual, $\pi_{\fr_l}$ and $\pi_\gr$ denote the maximally mixed states on $\fr_l$, respectively $\gr$ while $k_l:=\dim\fr_l$ gives the dimension of $\fr_l$.\\
For $j\in\{1,\ldots,N\}$, $\delta\in(0,1/2)$, $l\in\nn$ and states $\cn_j(\pi_{\gr})$ let $q_{j,\delta,l}\in \mathcal{B}(\kr)^{\otimes l}$ be the frequency-typical projection of $\cn_j(\pi_\gr)$ and $\cn_{j,\delta,l}$ be the reduced operation associated with $\cn_j$ and $\pi_\gr$ as given in Lemma \ref{lemma-typical-1}.\\
For an arbitrary unitary operation $u^l\in\mathcal B(\hr^{\otimes
l})$ we set
\begin{eqnarray*}
\hat\cn_{j,u^l,\delta}^l:=\mathcal{Q}_{j,\delta,l}\circ\cn_{j,\delta,l}\circ\mathcal
U^l,\ \ \
\hat\cn^l_{u^l,\delta}:=\frac{1}{N}\sum_{j=1}^{N}\hat\cn_{j,u^l,\delta}^l,\\
 \hat\cn_{j,\delta}^l:=\mathcal Q_{j,\delta,l}\circ\cn_{j,\delta,l},\ \ \ \ \hat\cn^l_\delta:=\frac{1}{N}\sum_{j=1}^{N}\hat\cn_{j,\delta}^l.
\end{eqnarray*}
Let $U^l$ be a random variable taking values in
$\mathfrak{U}(\gr^{\otimes l})$ which is distributed according to
the Haar measure. Application of Theorem \ref{convex-klesse}
yields
\begin{eqnarray}
\mathbb
E{F}_{c,e}(\pi_{\fr_l},\hat\cn^l_{U^l,\delta})&\geq&\textrm{tr}(\hat\cn^l_\delta(\pi_{\gr}^{\otimes
l}))\label{eq:direct-finite-uninformed-users-1}\\
&&-2\sum_{j=1}^{N}\sqrt{k_ln_{j,\delta,l}}||\hat\cn_{j,\delta}^l(\pi_{\gr}^{\otimes
l})||_2,\nonumber
\end{eqnarray}
where $n_{j,\delta,l}$ is the number of Kraus operators of
$\cn_{j,\delta,l}$. Notice that $\mathcal
Q_{j,\delta,l}\circ\cn_{j,\delta,l}$ has a Kraus representation
containing exactly $n_{j,\delta,l}$ elements. We will use
inequality (\ref{eq:direct-finite-uninformed-users-1}) in the
proof of the following Lemma.
\begin{lemma}[Direct Part for maximally mixed states]\label{lemma-direct-finite-uninformed-users-1}\
Let $\fri=\{\cn_1,...,\cn_N\}\subset \mathcal{C}(\hr,\kr)$ be a
compound channel and $\pi_\gr$ the maximally mixed state
associated to a subspace $\gr\subset\hr$. Then
$$Q(\fri)\geq\min_{\cn_i\in\fri}I_c(\pi_\gr,\cn_i).$$
\end{lemma}
\emph{Proof}. We show that for every $\epsilon>0$ the number $\min_{\cn_i\in\fri}I_c(\pi_\gr,\cn_i)-\epsilon$ is an achievable rate for $\fri$.\\
1) If $\min_{\cn_i\in\fri}I_c(\pi_\gr,\cn_i)-\epsilon\leq0$, there is nothing to prove.\\
2) Let $\min_{\cn_i\in\fri}I_c(\pi_\gr,\cn_i)-\epsilon>0$.\\
Choose $\delta\in(0,1/2)$ and $l_0\in\mathbb N$ satisfying $2\cdot\vphi(\delta)+h(l_0)<\epsilon/2$ with functions $\vphi,h$ from Lemma \ref{lemma-typical-1}.\\
For every $l\in\mathbb N$ let the dimension of the subspace
$\fr_l\subset\gr^{\otimes l}$ be given by
\[k_l=\lfloor2^{l(\min_{\cn_i\in\fri}I_c(\pi_\gr,\cn_i)-\epsilon)}\rfloor.\]
By $S(\pi_\gr)\geq I_c(\pi_\gr,\cn_j)$ (see \cite{bkn}), this is always possible.\\
We will now give lower bounds on the terms in
(\ref{eq:direct-finite-uninformed-users-1}), thereby making use of
Lemma \ref{lemma-typical-1}:
\begin{eqnarray}\label{eq:direct-finite-uninformed-users-2}
\textrm{tr}(\hat\cn^l_\delta(\pi_{\gr}^{\otimes l})
)\geq1-2\cdot2^{-l(c\delta^2-h(l))}.
\end{eqnarray}
A more detailed calculation can be found in \cite{bbn-1} or
\cite{klesse}. Further, using that
$||A+B||_2^2\geq||A||_2^2+||B||_2^2$ holds for nonnegative
operators $A,B\in\mathcal{B}(\kr^{\otimes l})$ (see
\cite{klesse}), we get the inequality
\begin{eqnarray}\label{eq:direct-finite-uninformed-users-3}
||\hat\cn_{j,\delta}^l(\pi_{\gr}^{\otimes l})||_2^2&\leq&2^{-l(S(\cn_j(\pi_{\gr}))-\vphi(\delta))}.
\end{eqnarray}
From (\ref{eq:direct-finite-uninformed-users-1}),
(\ref{eq:direct-finite-uninformed-users-2}),
(\ref{eq:direct-finite-uninformed-users-3}) and our specific
choice of $k_l$ we get for every $l\geq l_0$
\begin{eqnarray}
\mathbb
EF_{c,e}(\pi_{\fr_l},\hat\cn^l_{U^l,\delta})\geq1-2\cdot2^{-l(c\delta^2-h(l))}-2N\sqrt{2^{-l\epsilon/2}}\nonumber.
\end{eqnarray}
This shows the existence of at least one sequence $(\mathcal W^l,\crr^l)_{l\in\mathbb N}$ of
$(l,k_l)-$ entanglement transmission codes for $\fri$ and
\[ \liminf_{l\rightarrow\infty}\frac{1}{l}\log k_l=\min_{\cn_i\in\fri}I_c(\pi_{\gr},\cn_i)-\epsilon \]
as well as (using that entanglement fidelity is affine in the
channel), for every $l\in\mathbb N$ with $l\geq l_0$
\begin{equation} \min_{j\in\{1,\ldots,N\}}F_e(\pi_{\fr_l},\crr^l\circ\hat\cn_{j,\delta}^l\circ\mathcal W^l)\geq1-N\frac{1}{3}\epsilon_l \label{eq:direct-finite-uninformed-users-4}\end{equation}
where $\mathcal{W}^l(\cdot)=w^l(\cdot)w^{l\ast},\ w^l\in\mathfrak U(\gr^{\otimes l})\ \forall l\in\mathbb N$, and
\begin{equation}\label{eq:direct-finite-uninformed-users-eps}
\epsilon_l=3\cdot(2\cdot2^{-l(c\delta^2-h(l))}+2N\sqrt{2^{-l\epsilon/2}}).
\end{equation}
For every $j\in\{1,\ldots,N\}$ and $l\in\mathbb
N\backslash\{1,\ldots,l_0-1\}$ we thus have, by property 5) of Lemma \ref{lemma-typical-1}, construction of
$\hat\cn_{j,w^j,\delta}^l$, and equation
(\ref{eq:direct-finite-uninformed-users-4}),
\begin{eqnarray*}
&&F_e(\pi_{\fr_l},\crr^l\circ\mathcal Q_{j,\delta,l}\circ\cn_{j}^{\otimes l}\circ\mathcal W^l)\geq\\
&\geq&F_e(\pi_{\fr_l},\crr^l\circ\mathcal Q_{j,\delta,l}\circ\cn_{j,\delta,l}\circ\mathcal W^l)\\
&=&F_e(\pi_{\fr_l},\crr^l\circ\hat\cn_{j,w^j,\delta}^l)\\
&\geq&1-N\frac{1}{3}\epsilon_l.
\end{eqnarray*}
By Lemma \ref{gentle-operator-lemma-for-F_e}, this immediately
implies
\begin{equation*}
\min_{\cn_j\in\fri}F_e(\pi_{\fr_l},\crr^l\circ\cn_{j}^{\otimes
l}\circ\mathcal W^l)\geq1-N\epsilon_l\ \ \ \forall l\in\mathbb
N\backslash\{1,\ldots,l_0-1\}.
\end{equation*}
Since $\epsilon>0$ was arbitrary, we have shown that
$\min_{\cn_i\in\fri}I_c(\pi_{\gr},\cn_i)$ is an achievable rate. $\Box$
\\\\
For the proof of Theorem \ref{main-theorem} we only need one more
ingredient, which is a generalization of the well known BSST
Lemma of \cite{bsst}:
\begin{lemma}[Compound BSST Lemma, Cf. \cite{bbn-1}]\label{compound-bsst-lemma}
Let $\fri\subset\mathcal{C}(\hr,\kr) $ be an arbitrary set of
channels. For any $\rho\in \mathcal{S}(\hr)$ let
$q_{\delta,l}\in\mathcal{B}(\hr^{\otimes l})$ be the
frequency-typical projection of $\rho$ and set
\[ \pi_{\delta,l}:=\frac{q_{\delta,l}}{\textrm{tr}(q_{\delta,l})}\in\mathcal{S}(\hr^{\otimes l}). \]
Then there is a positive sequence $(\delta_l)_{l\in\nn}$
satisfying $\lim_{l\to\infty}\delta_l=0$ with
\[ \lim_{l\to\infty}\frac{1}{l}\inf_{\cn\in \fri}I_c(\pi_{\delta_l,l},\cn^{\otimes l} )=\inf_{\cn\in\fri}I_c(\rho,\cn).  \]
\end{lemma}\ \\
From Lemma \ref{lemma-direct-finite-uninformed-users-1} and the fact that
\begin{equation}Q(\fri^{\otimes l})=lQ(\fri)\label{eq:main-theorem-3}\end{equation} holds for every $l\in\mathbb N$ we
get independent from the value of $l$ and for every maximally mixed state $\pi_{\fr_l}\in\mathcal S(\hr^{\otimes l})$ supported on a subspace $\fr_l\subset\hr^{\otimes l}$ the inequality
\begin{equation}Q(\fri)\geq\frac{1}{l}\min_{\cn_i\in\fri}I_c(\pi_{\fr_l},\cn_i^{\otimes l}).\label{eq:main-theorem-1}\end{equation}
Let $\rho\in\mathcal S(\hr)$ be arbitrary and $(\delta_l)_{l\in\nn}$, $(\pi_{\delta_l,l})_{l\in\mathbb N}$ as in Lemma \ref{compound-bsst-lemma}. Then by (\ref{eq:main-theorem-1}) and Lemma \ref{compound-bsst-lemma} we have
\begin{eqnarray}Q(\fri)&\geq&\lim_{l\rightarrow\infty}\frac{1}{l}\min_{\cn_i\in\fri}I_c(\pi_{\delta_l,l},\cn_i^{\otimes l})\nonumber\\
&=&\min_{\cn_i\in\fri}I_c(\rho,\cn_i)\nonumber\\
&=&\min_{\cn_i\in\fri}I_c(\rho,\cn_i).\label{eq:main-theorem-2}
\end{eqnarray}
Thus, $Q(\fri)\geq\max_{\rho\in\mathcal S(\hr)}\min_{\cn_i\in\fri}I_c(\rho,\cn_i)$ has to hold. A second application of equation (\ref{eq:main-theorem-3}) and taking the limit $l\rightarrow\infty$ yields the desired result. $\Box$


\end{document}